\documentclass[%
reprint,
superscriptaddress,
 amsmath,amssymb,
 aps,
prb,
]{revtex4-2}

\usepackage{graphicx}
\usepackage{dcolumn}
\usepackage{bm}
\usepackage{siunitx}
\setcitestyle{super}

\newcommand*{\citen}{}
\DeclareRobustCommand*{\citen}[1]{%
  \begingroup
    \romannumeral-`\x 
    \setcitestyle{numbers}%
    \cite{#1}%
  \endgroup
}

\usepackage{xcolor}
\usepackage{soul}

\soulregister\cite7
\soulregister\citen7
\soulregister\ref7
\soulregister\textbf7
\soulregister\textit7

\usepackage{pdfpages}
\usepackage{pgffor}

\makeatletter
\AtBeginDocument{\let\LS@rot\@undefined}
\makeatother

\begin{document}

\preprint{APS/123-QED}
\title{Anisotropic magnon spin transport in CrPS\textsubscript{4}}

\author{K. Sundararajan}
\email{k.sundararajan@rug.nl}

\author{M. Zohaib}

\author{Y. Kreminska}
\author{S.H. Tirion}%

\author{Bart J. van Wees}
\affiliation{
Zernike Institute for Advanced Materials, University of Groningen, 9747 AG Groningen, The Netherlands}


 

\begin{abstract}

Crystal anisotropy provides a powerful route for realizing direction-dependent transport in solid-state systems. While its influence on electronic transport is well established, the role of anisotropy in magnon spin transport in van der Waals magnets is largely unexplored. Here, in a nonlocal geometry, utilizing the monoclinic van der Waals antiferromagnet CrPS\textsubscript{4}, we observe pronounced anisotropy in both electrically and in thermally excited magnon spin transport. Electrically generated magnons exhibit a magnon spin conductivity at least 2.2 times larger and a spin diffusion length at least 2.7 times longer for transport along the crystallographic-\textit{b} axis compared to the crystallographic-\textit{a} axis, where $\lambda_m^{a} \sim$ 211 nm and $\lambda_m^{b} \geq$ 575 nm. In comparison, at 8T, we find the nonlocal second-harmonic resistance associated with thermally excited magnons to be $\sim$7 times larger along the crystallographic-\textit{b} axis at 25K. We further show that a magnon spin diffusion length cannot be reliably extracted from the nonlocal second-harmonic resistance, owing to the extended temperature profile within CrPS$_4$. Likewise, we show that the anisotropy in the spin Seebeck coefficients cannot be reliably estimated from the thermally excited magnon spin transport alone, as it is intertwined with the anisotropic heat conductivity of CrPS$_4$. Utilizing the electrically generated magnon spin transport, we demonstrate that intrinsic crystalline anisotropy serves as an effective control parameter for tuning magnon spin transport, opening new avenues for magnonic device engineering.
\end{abstract}
\maketitle

\section{Introduction}

The study of collective excitations of magnetic order, namely spin waves and their quanta magnons has attracted significant interest due to the potential of magnon-based spin currents for low-power and high-speed information processing \citen{Magnon_Roadmap_1,Magnon_Roadmap_2}. Electrically controlled magnon spin transport has been demonstrated in a variety of magnetic systems \citen{Ludo_Nat,Wei_Nat, All_Electrical}, including ferrimagnetic \citen{Ludo_Nat,Wei_Nat} and antiferromagnetic oxides \citen{A2,A1,A3}, as well as two-dimensional (2D) van der Waals magnets \citen{Bart_CPS_1,Bart_CPS_2,Bart_CPS_3}. Two-dimensional van der Waals magnets provide a versatile platform for exploring fundamental magnon transport phenomena owing to their tunable magnetic properties \citen{Tunable1,Tunable2}, reduced dimensionality \citen{Nanometer1}, and strong magnetic anisotropy. In particular anisotropic exchange interactions can strongly influence magnon propagation and relaxation, leading to anisotropic magnon spin transport along different crystallographic axes \citen{Aniso_Ferrite}. Understanding such anisotropic magnon transport is therefore essential for the development of directional and energy-efficient magnonic devices.
\\

Anisotropic magnon spin transport has previously been observed in ferrimagnetic spinel ferrite thin films \citen{Aniso_Ferrite} and antiferromagnetic orthoferrites \citen{Klaui_Ani}, where the anisotropic transport of electrically excited magnons was attributed to wavevector-dependent exchange stiffness and direction-dependent magnon group velocities, respectively. Recently, anisotropic magnon transport has been theoretically discussed in van der Waals materials \citen{AdvMat_Aniso_Th} as well as in doped MXenes \citen{MX_Aniso}. Among the various van der Waals magnets, CrPS$_4$ has emerged as a particularly attractive platform for investigating anisotropic responses, in particular, anisotropy in optical responses \citen{CPS_OpticalAniso,CPS_OpticalAniso2}, in mechanical responses \citen{CPS_Mechanical}, in magnetic susceptibility \citen{SQUID_Ani} and in transport of thermally driven magnons \citen{NatCom_Aniso} have already been reported.\\

Magnon spin transport in CrPS\textsubscript{4} has been investigated extensively using electrically generated magnons \citen{Bart_CPS_1,Bart_CPS_2,Bart_CPS_3}, thermally generated magnons \citen{Bart_CPS_1,Bart_CPS_2,APL_ThermalMag,NatCom_Aniso,Klaui_CPS}, and antiferromagnetic resonance \citen{Coherent_CPS_2,Coherent_CPS_3}. In particular, Ref. \citen{NatCom_Aniso} reported anisotropic diffusion of thermally excited magnons in CrPS\textsubscript{4}. However, the influence of anisotropic magnetic exchange on magnon spin transport excited and detected electrically has not yet been investigated. Experimental studies of anisotropic magnon spin transport in van der Waals magnetic materials, furthermore have so far been largely limited to thermally generated magnons \citen{Aniso_Felix,NatCom_Aniso,NatCom_Crocl}.\\

In conventional vertical heavy-metal/magnetic-insulator/heavy-metal heterostructures, the temperature gradient across the magnetic layer is approximately uniform, generating a magnon spin current that produces a nonequilibrium magnon spin accumulation (or depletion) at the interface. This interfacial spin accumulation is detected electrically via the inverse spin Hall effect in the heavy-metal layer \citen{Vertical1}. In contrast, lateral nonlocal devices \citen{Ludo_Nat} exhibit inherently nonuniform temperature gradients generated by Joule heating. These spatially extended thermal gradients drive magnon spin currents over a finite region of the magnetic insulator, such that the measured nonlocal spin Seebeck response is determined not only by magnon transport between the injector and the detector but also by the complete temperature profile between them. Consequently, in the lateral nonlocal SSE geometry, the nonuniform temperature gradient generates magnon spin currents throughout the magnetic material. These thermally generated magnons diffuse towards the detector which is attenuated by the magnon spin diffusion length, $\lambda_m$. As a result, the measured nonlocal SSE response is a convolution of the spatial temperature profile and the magnon spin diffusion length. The challenges associated with extracting magnon spin diffusion lengths from nonlocal spin Seebeck measurements have been highlighted and have been discussed in detail previously in Refs. \citen{Bart_Thermal,Cassanova_Thermal,TwoOmegaYIG,gao2022magnon}. \\

\begin{figure*}[htb]
  \centering
  \includegraphics[width=\textwidth]{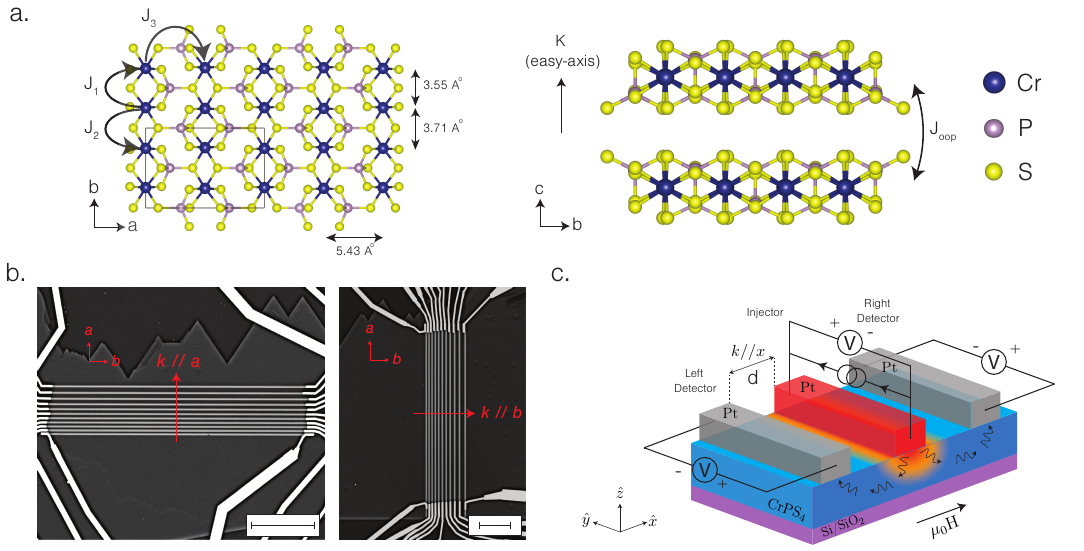}   
    \caption{a) Crystal structure of CrPS$4$. (Left) View in the \textit{ab} plane showing the nearest-neighbor Heisenberg exchange interactions, with $J_1$ and $J_2$ along the crystallographic \textit{b-} axis and $J_3$ along the \textit{a-} axis. (Right) View in the \textit{bc} plane illustrating the interlayer exchange coupling $J{_\mathrm{oop}}$ and the bulk easy axis $K$ along the crystallographic \textit{c} axis. (b) Scanning electron micrographs of device S1 with Pt injector and detector electrodes aligned along the crystallographic \textit{a-} (k$\parallel$b) and \textit{b-} (k$\parallel$a) axes of CrPS$_4$. Scale bars: 10 $\mu$m. (c) Schematic illustration of the nonlocal device geometry used to investigate magnon spin transport in CrPS$_4$, with the sign convention used for the measurements. The magnon propagation direction is denoted by $k \parallel \hat{x}$, where $\hat{x}$ is oriented along either the crystallographic \textit{a} or \textit{b} axis.}
    \label{fig:Intro_Pic}
\end{figure*}

In addition, the temperature dependence of the thermal conductivity alters the spatial distribution of the temperature profile, affecting both the thermally driven magnon spin current and the associated transport coefficients. As we show here, for CrPS$_4$, the various mechanisms involved in the generation and detection of thermally excited magnons render the extraction of the magnon spin diffusion length from the nonlocal second-harmonic resistance, attributed to the nonlocal spin Seebeck effect inaccurate and prone to overestimation. Furthermore, the anisotropy in the magnitude of the nonlocal second-harmonic resistance is further intertwined with the anisotropy of the heat conductivity, making it difficult to attribute the observed anisotropy to the nonlocal spin Seebeck coefficients alone. \\

We note that electrical magnon spin injection and detection provides a well-defined transport channel compared to the detection of thermally excited magnons, enabling a direct characterization of magnon spin transport and thus allows us to isolate the effects of exchange anisotropy on the magnon spin transport in CrPS$_4$. In the lateral nonlocal geometry, a nonequilibrium magnon spin accumulation (or depletion) is generated locally beneath the injector electrode enabled by the spin Hall effect (SHE). The resulting magnon chemical potential gradient drives magnon diffusion towards the detector electrode, wherein the magnon spin accumulation is converted into a measurable open-circuit voltage via the inverse SHE \citen{Ludo_Nat}. Consequently, the electrically generated magnon transport enables defining an unambiguous transport distance set by the injector–detector separation. Using electrically excited magnons, at 25K, we find an anisotropy in the magnon spin conductivity of at least a factor of 2.2 and a spin diffusion length at least 2.7 times longer for transport along the crystallographic $b-$ axis compared to the crystallographic $a-$ axis.\\


\section{Results and Discussion}

\subsection*{Magnetic structure of CrPS$_4$}

To study the effect of anisotropic exchange coupling on magnon spin transport, we use CrPS\textsubscript{4}, an A-type antiferromagnetic (charge-)insulator, as the magnon transport medium. CrPS\textsubscript{4} crystallizes in a monoclinic structure wherein the Cr$^{3+}$ ions couple ferromagnetically within each layer and antiferromagnetically between adjacent layers \citen{SQUID_CPS,SQUID_Ani}. The crystal structure of CrPS\textsubscript{4} lacks in-plane four-fold rotational symmetry \citen{AdvMat_Aniso_Th}, as illustrated in Fig. \ref{fig:Intro_Pic}a, which allows for anisotropic exchange interactions, which have been confirmed by inelastic neutron scattering experiments \citen{CPS_Neutron}. Since magnon spin dynamics are governed predominantly by spin–spin interactions described by the Heisenberg exchange interaction, CrPS\textsubscript{4} provides an ideal system for exploring how exchange anisotropy influences magnon-mediated spin transport.\\

The Hamiltonian that describes the magnetic ground state of CrPS$_4$ under an applied external magnetic field ($B$) below the N\'eel temperature ($\sim$38~K for bulk CrPS$_4$ \citen{SQUID_CPS}) is given by: 

\begin{equation}
    \begin{aligned}
    \mathcal{H} = &\sum_{ i,j } J_{ij} \vec{S}_i\cdot\vec{S}_j + \sum_{i} K_z(\vec{S}_i \cdot \hat{z})^2
    - \sum_i\mu_i\vec{B}\cdot\vec{S}_i, \\
\end{aligned}
\end{equation}

where $J$ denotes the exchange interactions and $K$ denotes the single-ion anisotropy, which favors the alignment of the sublattice magnetization along the out-of-plane direction. As shown in Fig. \ref{fig:Intro_Pic}a, CrPS\textsubscript{4} exhibits strong in-plane anisotropic exchange coupling with exchange constants $J_{1} = -2.96$ meV, $J_{2} = -2.09$ meV, and $J_{3} = -0.51$ meV \citen{CPS_Neutron}. In contrast, the out-of-plane antiferromagnetic exchange coupling and single-ion anisotropy are much weaker, with $J_{\mathrm{oop}} = +0.16$ meV and $K = -0.0058$ meV. Owing to the weak out-of-plane exchange coupling and single-ion anisotropy, CrPS\textsubscript{4} for an out-of-plane magnetic field undergoes a spin-flop transition at approximately 0.8 T and reaches a fully saturated collinear magnetic state around 8 T \citen{SQUID_CPS,SQUID_Ani}. For magnetic fields applied in the plane, perpendicular to the easy axis, no spin-flop transition occurs. Instead, the applied magnetic field causes a continuous canting of the spins towards the magnetic field and eventually results in a fully collinear state at fields of approximately 8~T \citen{SQUID_CPS,SQUID_Ani}.\\

It was shown in Ref. \citen{Ludo_Eq} that for magnons with a quadratic dispersion relation of the form $\hbar\omega_k = J_S k^2 + \Delta$, where $J_S$ is the exchange stiffness, $k$ is the wave vector, and $\Delta$ is the spin-wave gap, the magnon spin diffusion length scales as $\lambda_m \propto \sqrt{J_S}$, while the magnon spin conductivity scales linearly with the exchange stiffness, i.e., $\sigma_m \propto J_S$. Although Ref.~\citen{Ludo_Eq} focuses on the ferrimagnetic insulator YIG, for which the spin-wave gap is negligible, it establishes that the exchange stiffness is a key parameter governing diffusive magnon spin transport governing both $\lambda_m$ and $\sigma_m$. Since the exchange stiffness reflects the strength of the microscopic exchange interactions that stabilize the magnetic order and is approximately proportional to the relevant exchange couplings~\citen{rezende2020fundamentals}, CrPS$_4$, with its pronounced in-plane exchange anisotropy between the crystallographic \textit{a-} and \textit{b-} axes, provides an ideal platform for investigating anisotropic magnon spin transport in van der Waals magnets.
\\ 

\subsection*{Measurement Geometry}

We employ a nonlocal magnon transport geometry \citen{Ludo_Nat} to study the effect of exchange anisotropy on the magnon spin transport in CrPS$_4$. One of the measurement devices used in this work is shown in Fig.~\ref{fig:Intro_Pic}b, while Fig.~\ref{fig:Intro_Pic}c illustrates the nonlocal measurement geometry. The device consists of $\sim$10 nm thick platinum (Pt) strips, approximately 500 nm wide with various edge-to-edge distances ($d$) between them, fabricated on top of CrPS\textsubscript{4} (see Supplemental Material I for an overview of the various device parameters). For this study, all devices were fabricated with the Pt strips oriented along either the crystallographic \textit{a}- or \textit{b}-axis of CrPS\textsubscript{4} (the small deviations in the orientation of the platinum electrodes with respect to the crystallographic axes are summarized in Table I of Supplementary Information for the various devices measured). The principal crystallographic directions of CrPS$_4$ were determined following Ref. \citen{SI_Exfo} (see Supplemental Material I for details). To probe the magnon spin transport along the crystallographic-\textit{a} axis, the platinum strips were patterned with their long axis parallel to the \textit{b-} crystallographic axis and is denoted as $k$$\parallel$a (and $k$$\parallel$b for strips oriented along the \textit{a-} axis), where $k$ denotes the magnon propagation direction.\\

For the electrical measurements, an alternating charge current, $i_{ac}=\sqrt{2}\, i_0 \sin(\omega t)$ (\textit{f} = 17.777 Hz unless mentioned otherwise), is sent through the Pt injector electrode. The charge current in the injector platinum electrode generates a perpendicular spin current due to the spin Hall effect that results in a spin accumulation at the Pt/CrPS$_4$ interface. When the polarization of this resulting spin accumulation is parallel (antiparallel) to the magnetization of CrPS$_4$, magnons are annihilated (created), producing a nonequilibrium magnon accumulation that diffuses through CrPS$_4$ and is electrically detected at the Pt detector electrode via the inverse spin Hall effect \citen{Ludo_Nat}.\\

In addition to the electrically generated and detected magnons, Joule heating of the injector ($\propto i^2R$) locally increases the temperature of the magnetic insulator, thereby establishing a temperature gradient that in turn drives a heat current through CrPS$_4$. This heat current drives a pure magnonic spin current, unaccompanied by any charge current. A magnon spin accumulation therefore builds up at the boundaries of CrPS$_4$, which can induce a flow of spin angular momentum into an adjacent platinum (Pt) electrode, where it is converted into a measurable electrical voltage by the inverse spin Hall effect (iSHE). Depending on the device geometry, two distinct measurement configurations can be realized. When the thermal magnon excitation and iSHE detection occur at the same Pt electrode, the measured response is referred to as the local spin Seebeck effect (SSE). In contrast, when the heater and detector electrodes are spatially separated, the measured response is referred to as the nonlocal SSE. As already discussed, since heat transport is diffusive, the temperature gradient, $\nabla T \equiv \nabla T(x)$, extends over a substantial region of the magnetic insulator, driving thermally excited magnon spin currents far beyond the injector. Consequently, the nonlocal SSE depends not only on the magnon spin diffusion through the magnetic insulator but also on the complete spatial temperature distribution.\\

For an ac current through the injector platinum electrode, we define the nonlocal first- and second-harmonic resistances measured at the detector electrode as $R^{\mathrm{NL}}_{1\omega}=V^{\mathrm{NL}}_{1\omega}/i_0$ and
$R^{\mathrm{NL}}_{2\omega}=V^{\mathrm{NL}}_{2\omega}/i_0^2$, respectively. In the present measurement geometry, $R^{\mathrm{NL}}_{1\omega}$ originates from electrically generated and detected magnon spin transport, whereas $R^{\mathrm{NL}}_{2\omega}$ arises from the nonlocal SSE associated with thermally driven magnon spin current. We attribute the observed nonlocal second-harmonic resistance to the nonlocal spin Seebeck effect (SSE) and rule out other thermoelectric contributions (see Supplemental Material III for details ~\citen{Sup_Inf}). Analogously, $R^{\mathrm{Loc}}_{2\omega}$ is attributed to the local SSE. The nonlocal transport measurements were performed by simultaneously measuring the first- and second-harmonic resistances at the injector and detector electrodes. For all measurements reported here, the in-plane magnetic field was applied perpendicular to the platinum electrodes. The relative orientation of the magnetic field with respect to the platinum electrodes was determined from the angular dependence of the local and nonlocal second-harmonic resistances measured as a function of the in-plane magnetic field angle (see Supplemental Material III for details ~\citen{Sup_Inf}). We note that all second-harmonic resistances are anti-symmetrized and the first-harmonic resistances are symmetrized with respect to the magnetic field.\\

We organize the discussion as follows. We first examine the local second-harmonic resistance, $R^{\mathrm{Loc}}_{2\omega}$, attributed to the local spin Seebeck effect (SSE), and discuss the influence of the exchange anisotropy on the measured response. We then consider the nonlocal first-harmonic resistance, $R^{\mathrm{NL}}_{1\omega}$, arising from electrically injected and detected magnons, focusing on the anisotropy of magnon spin transport along the crystallographic \textit{a-} and \textit{b-} axes. Finally, we discuss the nonlocal second-harmonic resistance, $R^{\mathrm{NL}}_{2\omega}$, attributed to the nonlocal SSE. We then proceed to analyze the distance dependence of $R^{\mathrm{NL}}_{2\omega}$ along the two crystallographic axes. We further show that since the nonlocal SSE is governed by the temperature gradient, the pronounced anisotropy in the thermal conductivity further complicates the isolation of the intrinsic anisotropy of the SSE from the measured nonlocal second-harmonic response.
\subsection*{Local second-harmonic response}

Since the local first-harmonic response of the Pt strips, attributed to spin Hall magnetoresistance (SMR), is expected to be strongly influenced by the interfacial magnetic structure~\citen{Felix_SMR,Bart_CPS_3}, which may differ substantially from the bulk magnetic properties of CrPS$_4$, we do not focus on the SMR response in this work. A detailed discussion of this is provided in Supplemental Material III~\citen{Sup_Inf}.\\

\begin{figure}[h]
\centering
\includegraphics[width=0.85\linewidth]{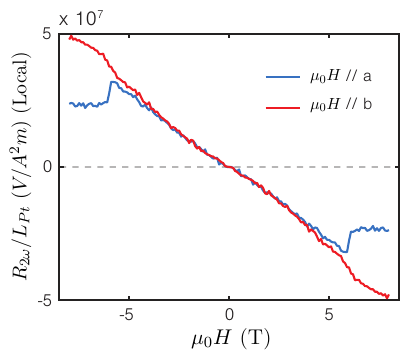}    \caption{\label{fig:LocSH} Magnetic field dependence of the local second-harmonic resistance (normalized by the Pt electrode length) along the different crystallographic axes of Device S1 (T = 25K, i$_{0}$ = 75 $\mu$A).}
\end{figure}

Fig. \ref{fig:LocSH} shows the local second-harmonic resistance at 25K, attributed to the local SSE, for the injector electrodes oriented along the crystallographic \textit{a}- and \textit{b}-axes as a function of the in-plane applied magnetic field. We observe that upto the spin-flip transition (around 6T), the magnitude the local second-harmonic resistance (normalized by the electrode length) exhibits a similar field dependence for both orientations. For Pt electrodes oriented along the crystallographic \textit{a-} axis, the dominant thermal gradients are $|\nabla_b T|$ and $|\nabla_c T|$, whereas for Pt electrodes oriented along the crystallographic \textit{b-} axis they are $|\nabla_a T|$ and $|\nabla_c T|$. The observation that the local second-harmonic resistance exhibits a similar magnetic-field dependence for both electrode orientations up to the spin-flip transition suggests that magnon diffusion along the out-of-plane (\textit{$\hat{c}$}) direction, driven predominantly by the out-of-plane temperature gradient $|\nabla_c T|$, dominates the local spin Seebeck response. Above the spin-flip transition, however, the local SSE exhibits distinct magnetic-field dependences for the two crystallographic orientations, indicating the emergence of an additional contribution.
\\ 

A possible origin of the different magnetic field dependence of the local SSE is the onset of in-plane magnon spin transport above the spin-flip transition, as inferred from the nonlocal first-harmonic measurements seen in Fig. \ref{fig:NLFH} and is discussed in detail later. Above the spin-flip field, this additional transport channel contributes to magnon diffusion within the \textit{ab} plane of CrPS$_4$, therefore modifying the field dependence of the local SSE. Specifically, below the spin-flip transition, magnon diffusion away from the injector is predominantly driven by the out-of-plane thermal gradient, $|\nabla_cT|$. However, above the spin-flip transition, the in-plane thermal gradients, $|\nabla_aT|$ or $|\nabla_bT|$, also contribute, depending on the orientation of the injector electrode with respect to the crystallographic axes. The temperature dependence of the local second-harmonic resistance is presented in Supplemental Material VIII \citen{Sup_Inf}, where a systematic shift in the magnetic-field position associated with the lineshape modification of the local second-harmonic response is observed, following the temperature dependence of the spin-flip transition of CrPS$_4$. Finally, we emphasize that the observed magnetic field dependence of the local SSE is reproducible across different injector electrodes of Device S1 and exhibits qualitatively similar behavior in all investigated samples (see Supplemental Material XI ~\citen{Sup_Inf} for additional data).\\

We note that, in Fig.~\ref{fig:LocSH}, the measurements of the local SSE are performed by changing both the orientation of the Pt strips and the applied magnetic field. Furthermore, the observation of a comparable local second-harmonic resistance up to the spin-flip field indicates that the orientation of the magnetic field does not significantly influence the measured response, namely that $|\nabla_cT(B_a)|$ and $|\nabla_cT(B_b)|$ does not have a significant effect on the local SSE before the spin-flip transition. In the present measurement geometry, we further exclude a contribution from the ordinary Nernst effect to the local SSE signal, $V_y=\nu_{\mathrm{Pt}}\nabla_zT\,B_x$, where $\nu_{\mathrm{Pt}}$ is the Nernst coefficient of platinum, which arises from the combination of an out-of-plane thermal gradient and an in-plane magnetic field perpendicular to the Pt strip. A detailed discussion is provided in Supplemental Material III \citen{Sup_Inf}. The physical origin of the observed differences in the local second-harmonic response, after the spin-flip transition remains an open question.


\subsection*{Nonlocal first-harmonic response}

\begin{figure}[h]
\centering
\includegraphics[width=0.85\linewidth]{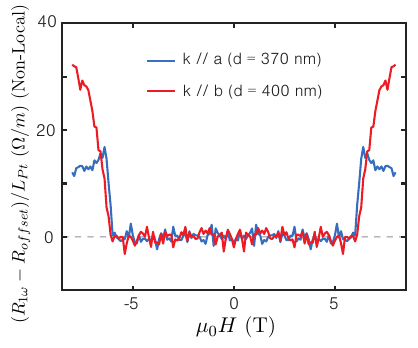}    \caption{\label{fig:NLFH} Magnetic field dependence of the nonlocal first-harmonic resistance (normalized by the Pt electrode length) along the different crystallographic axes of Device S1 (T = 25K, i$_{0}$ = 75 $\mu$A).}
\end{figure}

Fig.~\ref{fig:NLFH} shows the nonlocal first-harmonic resistance, $R^{\mathrm{NL}}_{1\omega}$, as a function of the applied magnetic field for magnon spin transport along the crystallographic \textit{a} and \textit{b} axes. The bias- and frequency-dependent offsets, likely caused by capacitive coupling in the measurement circuit in $R_{1\omega}^{\mathrm{NL}}$ have been subtracted to emphasize the magnetic field dependence of the nonlocal first-harmonic resistance. We note that the qualitative magnetic field dependence of the magnon spin transport is similar along both the crystallographic axes. In particular, the magnon spin transport is completely suppressed below the spin-flip transition of CrPS$_4$ as inferred from the absence of detectable nonlocal first-harmonic resistance up to the spin-flip field, followed by a rapid onset of the nonlocal resistance above the spin-flip field. This behavior is consistent with previous reports of magnon spin transport in CrPS$_4$ \citen{Bart_CPS_1, Bart_CPS_2}.\\

We observe that for comparable injector–detector separations, above $\sim$6T, R$^{NL}_{1\omega}$ arising from magnon spin transport is larger along the crystallographic \textit{-b} axis. Within the diffusive magnon transport framework \citen{Ludo_Eq,Wei_Nat}, the observed anisotropy originates from the combined contributions of the magnon spin conductivity ($\sigma_m$) and the magnon spin diffusion length ($\lambda_m$). To disentangle the anisotropy associated with these two parameters, we perform distance-dependent measurements of the nonlocal first-harmonic resistance as a function of magnetic field for different injector-detector separations ($d$).\\

Following Ref. \citen{Ludo_Nat,Wei_Nat}, the nonlocal first-harmonic resistance (normalized by the length of the electrodes, in $\Omega$/m) arising from diffusive magnon spin transport, under the assumption of a negligible interfacial spin resistance at the Pt/CrPS$_4$ interface is given by:

\begin{equation} \label{Eqn: Ballistic}
\begin{aligned}
R_{NL} &= \frac{\sigma_{m}\, t_{\mathrm{CPS}}\, \eta_{\mathrm{Pt}}^{2}}{\lambda_{m}}\, \operatorname{csch}\!\left( \frac{d}{\lambda_{m}} \right)\\ 
&= \frac{C}{\lambda_{m}}\, \operatorname{csch}\!\left( \frac{d}{\lambda_{m}} \right).
\end{aligned}
\end{equation}

where csch(x) = $\frac{2}{e^{x}-e^{-x}}$ and C = $\sigma_{m}\, t_{\mathrm{CPS}}\, \eta_{\mathrm{Pt}}^{2}$ and $\sigma_m$, t$_{CPS}$, $\lambda_m$ and $d$ correspond to the the magnon spin conductivity, thickness of CrPS$_4$, the magnon spin diffusion length and the edge-to-edge distance between the injector and the detector electrodes. $\eta_{Pt}$ characterizes the effective charge-to-spin interconversion in the platinum electrode (see Supplemental Material V \citen{Sup_Inf}). The nonlocal resistance described by Eq.~\ref{Eqn: Ballistic} exhibits two characteristic regimes determined by the injector--detector separation $d$ relative to the magnon spin diffusion length $\lambda_m$. In the short-distance limit ($d \ll \lambda_m$), magnon relaxation is negligible and the transport is ohmic wherein $R_{\mathrm{nl}} \propto 1/d$. Conversely, in the long-distance limit ($d \gg \lambda_m$), magnon relaxation dominates and results in an exponential decay of the nonlocal resistance with increasing separation \citen{Ludo_Nat}.\\ 

\begin{figure}[h]
\centering
\includegraphics[width=0.95\linewidth]{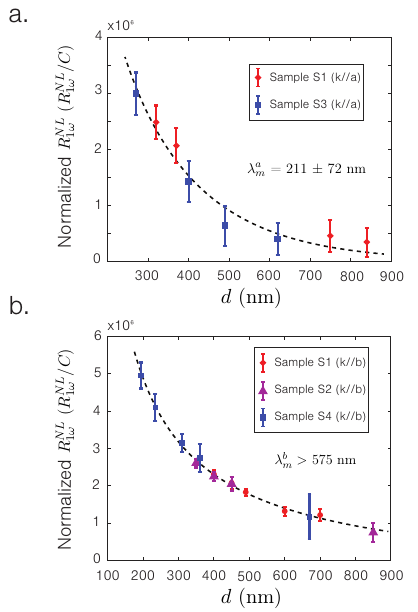}    \caption{\label{fig:Diffusion} The normalized nonlocal first-harmonic resistance ($R^{NL}_{1\omega}/(\sigma_{m} t_{CPS} \eta_{Pt}^{2})$) as a function of the edge-to-edge distance between the platinum strips for magnon spin diffusion along the crystallographic a) \textit{a-} and b) \textit{-b} axes (where $\lambda_m^b = $575 nm).}
\end{figure}

Fig. \ref{fig:Diffusion} a,b shows the distance dependence of the nonlocal first-harmonic resistance measured along the crystallographic \textit{a-} and \textit{-b} axes, respectively, from four different samples. To account for sample-to-sample variations in the amplitude of the nonlocal resistance arising from differences in the magnon spin conductivity and thickness of the platinum strips and CrPS$_4$, we normalize the nonlocal resistance by the sample-dependent pre-factor, namely $\mathrm{C}$ which depends on $\sigma_m t_{CPS} \eta_{Pt}^2$, and plot the normalized $R^{1\omega}_{NL}$ ($R^{1\omega}_{NL}/\mathrm{C}$), i.e., $\frac{\mathrm{csch}(d/\lambda_m)}{\lambda_m}$ (for details see Supplemental Material V \citen{Sup_Inf}). We observe that, for CrPS$_4$ thicknesses between 40 and 60~nm studied here, the decay of the nonlocal first-harmonic resistance as a function of $d$ along the different crystallographic axes is well described by a common characteristic decay length across all samples, suggesting that the magnon spin diffusion length is does not vary significantly for thicknesses of CrPS$_4$ within this range.
\\

From the distance dependence of the nonlocal first-harmonic resistance, we extract magnon spin diffusion lengths of $\lambda_m^{a} = 211 \pm 72$ nm and $\lambda_m^{b} \ge 575$ nm for the magnon spin transport along the principal crystallographic axes in CrPS$_4$. We observe that the distance dependence of the nonlocal first-harmonic resistance along the crystallographic \textit{-b} axis can also be described by a 1/d decay (for details see Supplemental Material V \citen{Sup_Inf}). This observation indicates that the transport along the crystallographic \textit{b-} axis occurs in the regime wherein the injector-to-detector separation is comparable to or smaller than the magnon spin diffusion length ($d\lesssim\lambda_m$). As a result, we emphasize that the value of $\lambda_m^{b}$ extracted from the fitting of the distance dependence of the nonlocal first-harmonic resistance along the crystallographic \textit{-b} axis to \ref{Eqn: Ballistic}, as discussed previously, should be regarded as a lower bound on the magnon spin diffusion length. Since the measured distances predominantly probes the regime wherein magnon relaxation is negligible, the fit is only weakly sensitive to $\lambda_m$, leading to a possible underestimate to the value of $\lambda_m^{b}$ (for details see Supplemental Material V \citen{Sup_Inf}).\\

For device S1 wherein we measure the anisotropic magnon spin transport response on the same CrPS$_4$ flake, reducing device to device variations, we observe that $\frac{\lambda^{b}_m}{\lambda^{a}_m} \sim 2.22 \pm 0.89$ (and $\frac{\lambda^{b}_m}{\lambda^{a}_m}  \sim 2.72 \pm 0.96$ when taking into account all the devices measured here as shown in Fig. \ref{fig:Diffusion}). Similarly, for device S1, we find that the magnon spin conductivity is anisotropic wherein the extracted magnon spin conductivities are $\sigma_m^{a} = (1.24 \pm 0.40) \times 10^{5}~\mathrm{S,m^{-1}}$ and $\sigma_m^{b} = (2.76 \pm 0.23) \times 10^{5}~\mathrm{S,m^{-1}}$, and thus yielding $\frac{\sigma^{b}_m}{\sigma^{a}_m} \sim 2.22 \pm 0.90$. We note that along the crystallographic \textit{-b} axis, the nonlocal first-harmonic resistance does not saturate even at an applied magnetic field of 8 T, as shown in Fig. \ref{fig:NLFH}. Consequently, the reported value of $\sigma_m^{b}$ is underestimated, despite already exceeding $\sigma_m^{a}$. The extracted magnon spin conductivities and spin diffusion lengths from all the devices used are summarized in Table S3 in the Supplemental Material V \citen{Sup_Inf}. From the nonlocal first-harmonic resistance, we observe that the magnon spin transport exhibits pronounced anisotropy, with both the magnon spin conductivity and the magnon spin diffusion length being larger along the crystallographic \textit{b}-axis in comparison to that along the \textit{a}-axis.\\

We note that the previously reported anisotropic magnon spin transport in Refs.~\citen{Aniso_Ferrite,Klaui_Ani} was observed in systems with an easy-axis magnetocrystalline anisotropy lying within the magnon propagation plane. In contrast, for CrPS$_4$, the easy axis is oriented along the crystallographic \textit{c-} axis, such that the magnon wave vector ($\mathbf{k}\parallel\textit{a}$ or $\mathbf{k}\parallel\textit{b}$) remains perpendicular to the easy axis in our measurements. The observed anisotropy in magnon spin transport therefore cannot be attributed to the orientation of the easy axis with respect to the propagation direction, but instead arises directly from the anisotropic exchange coupling. Furthermore, the magnon spin transport in CrPS$_4$ is observed only in the collinear magnetic state above the spin-flip transition which indicates that magnon spin transport is governed by the field-induced magnetization rather than the N\'eel vector, in contrast to the behavior reported for the orthoferrite YFeO$_3$ in Ref.~\citen{Klaui_Ani}.\\

The observation of anisotropic magnon spin conductivity in the magnetic insulator CrPS$_4$ is expected to give rise to transport phenomena analogous to those resulting from anisotropic charge transport. In anisotropic electrical conductors \citen{ivchenko2021coulomb}, an intrinsic anisotropic resistivity tensor generates a transverse voltage when the electrical charge current is oriented along a direction that is not aligned with a principal crystallographic axis. Similarly, when the magnon chemical potential gradient is oriented away from the principal crystallographic axes, an anisotropic magnon spin conductivity tensor generates both longitudinal and transverse magnon spin currents, as described by $\mathbf{j}_m=-\mathbf{\sigma}_m\nabla\mu_m/e$. For a magnon chemical potential gradient, $\nabla\mu_m$, oriented at an angle $\theta$ with respect to the crystallographic \textit{a}-axis, the relation between the magnon spin current and the magnon chemical potential gradient given by the magnon spin conductivity tensor can be expressed as:

\begin{equation}
\label{eq:magnon PHE}
    \mathbf{\sigma_m}(\theta)=\left(\begin{array}{cc}
\sigma_m^a \cos ^2 \theta+\sigma_m^b \sin ^2 \theta & \left(\sigma_m^a-\sigma_m^b\right) \sin \theta \cos \theta \\
\left(\sigma_m^a-\sigma_m^b\right) \sin \theta \cos \theta & \sigma_m^a \sin ^2 \theta+\sigma_m^b \cos ^2 \theta
\end{array}\right),
\end{equation}

where $\mathbf{\sigma_m}(\theta)$, $\sigma_m^{a/b}$ and $\theta$ correspond to the magnon spin conductivity tensor, the magnon spin conductivity along the principal crystallographic axes and the angle between the propagation direction and the crystallographic \textit{a-} axis (for details see Supplemental Material VII \citen{Sup_Inf}).\\ 

This generation of a transverse magnon spin current in response to a longitudinal gradient in the magnon spin chemical potential originates entirely from the intrinsic crystallographic anisotropy of the magnon spin conductivity tensor, in contrast to the conventional magnon Hall effect \citen{onose2010observation}, where the transverse response is driven by Dzyaloshinskii–Moriya interaction. Furthermore, this mechanism is fundamentally different from the magnon planar Hall effect (MPHE) reported in Ref.~\citen{Magnon_PHE}, wherein the anisotropy originates from the dependence of the magnon spin conductivity on the relative orientation between the magnon spin current and the magnetization, directly analogous to electronic anisotropic magnetoresistance and the planar Hall effect. Although a similar MPHE could exist in CrPS$_4$, the MPHE-like effect proposed here arises from the conductivity anisotropy determined entirely by the crystal lattice and does not inherently rely on the orientation of the magnetization. It is therefore more closely analogous to the transverse Seebeck effect in anisotropic crystals, where an off-axis thermal gradient produces a transverse thermoelectric voltage through the intrinsic anisotropy of the Seebeck tensor \citen{Uchida_Trans_Seebeck}. The observation of anisotropic magnon spin conductivity makes CrPS$_4$ a promising platform for exploring the MPHE-like effects within the family of magnetic van der Waals materials.\\

Similarly, the magnon spin diffusion length for propagation at an arbitrary angle $\theta$ is then given by: 

\begin{equation}
\label{eqn:lambda}
\lambda_m^2(\theta) = (\lambda_m^a)^2\cos^2\theta + (\lambda_m^b)\sin^2\theta,
\end{equation}

where $\lambda_m(\theta)$ and $\lambda_m^{a/b}$ denote the magnon spin diffusion along $\hat{k} = (\cos\theta,\sin\theta)$ and the principle crystallographic axes respectively. For magnon spin transport along directions that are oriented away from the principal crystallographic axes, the magnon spin diffusion length changes continuously with the propagation direction (for details see Supplemental Material VII \citen{Sup_Inf}).
\subsection*{Nonlocal second-harmonic response}

\begin{figure}[h]
\centering
\includegraphics[width=0.85\linewidth]{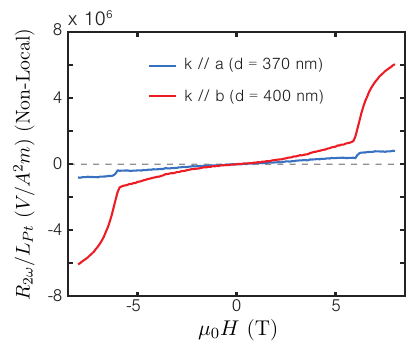}    \caption{\label{fig:NLSH} Magnetic field dependence of the nonlocal second-harmonic resistance (normalized by the Pt electrode length) along the different crystallographic axes of Device S1 (T = 25K, i$_{0}$ = 75 $\mu$A).}
\end{figure}

Fig. \ref{fig:NLSH} shows the nonlocal second-harmonic resistance, $R^{\mathrm{NL}}_{2\omega}$, attributed to the nonlocal SSE, as a function of the applied magnetic field for magnon spin transport along the crystallographic \textit{a-} and \textit{b-} axes at 25K. Analogous to the trend observed in Fig. \ref{fig:NLFH}, we observe a rapid onset in $R^{\mathrm{NL}}_{2\omega}$ above the spin-flip transition of CrPS$_4$, suggesting that the increase in $R^{\mathrm{NL}}_{2\omega}$ is related to the in-plane magnon spin transport. Similar to the trend observed for $R^{\mathrm{NL}}_{1\omega}$, we observe that for comparable injector–detector separations, $R^{\mathrm{NL}}_{2\omega}$ is larger along the crystallographic \textit{b-} axis. However, we observe that the magnitude of the anisotropy in the nonlocal first- and second-harmonic resistances differs significantly, with the resistances along the two crystallographic axes differing by factors of $\sim$2 and $\sim$7 respectively at 8T. Furthermore, we observe that the anisotropy in the nonlocal second-harmonic resistance is magnetic-field dependent and is observed even below the spin-flip field, where in-plane magnon spin transport is suppressed. This suggests that distinct mechanisms govern electrically generated and detected magnons and thermally driven magnons. Furthermore, apart from the anisotropy in the nonlocal second-harmonic resistance along the different crystallographic axes, we note that its magnetic-field dependence exhibits distinct line shapes for the two transport directions, the physical origin of which remains unclear. The bias and frequency dependence of the nonlocal first- and second-harmonic resistances are presented in Supplemental Material IV, while the temperature dependence of the nonlocal first- and second-harmonic resistances are presented in Supplemental Materials IX, and X \citen{Sup_Inf}.\\


As already discussed, we emphasize that, in the case of nonlocal SSE, the thermally excited magnon spin transport is intrinsically intertwined with several other parameters beyond magnon spin transport alone. In particular, a thermal magnon spin current is excited wherever a temperature gradient ($\nabla$T) exists, making the excitation process highly nonlocal and strongly dependent on the full temperature landscape within the magnetic medium. Unlike electrically injected magnons, the thermally generated magnon population is governed not only by the magnon diffusion length ($\lambda_m$), but also by the spatial configuration of the temperature profile. This issue is further complicated by the fact that the nonlocal magnon spin transport measurements reported here are performed on SiO$_x$/Si substrates. At low temperatures, the thermal conductivity of doped-Si is substantially larger than that of the GGG substrates typically used for YIG, leading to a markedly different distribution of heat flow and temperature profile \citen{Kappa_Si,Kappa_GGG,euler2015thermal}. Consequently, the spatial profile of thermally generated magnons can differ significantly from that in YIG/GGG devices, further complicating the interpretation of the nonlocal spin Seebeck measurements reported here. It is therefore crucial to understand the influence of SiO$_x$ and doped-Si on the temperature profile in thermally driven magnon spin transport studies of van der Waals magnets.\\

\begin{figure*}[htb]
  \centering
  \includegraphics[width=\textwidth]{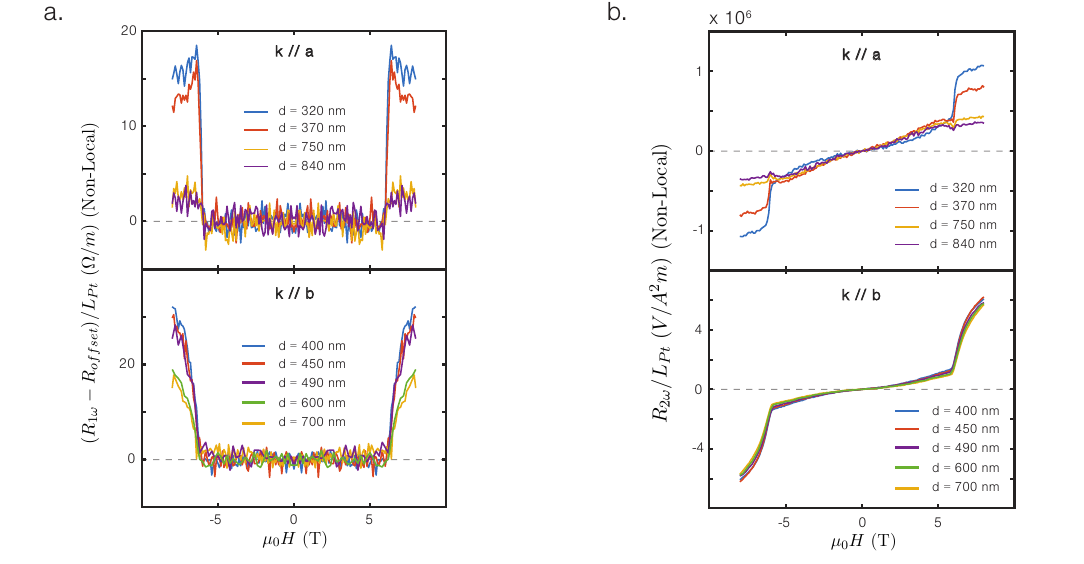}   
    \caption{The magnetic field dependence of a) the nonlocal first-harmonic resistances and b) nonlocal second-harmonic resistances normalized by the pt electrode length for various distances between the injector and detector electrodes of Device S1 (T = 25K, i$_{0}$ = 75 $\mu$A).}
    \label{fig:Distance_dependence}
\end{figure*}

Consequently, an extended and spatially inhomogeneous temperature profile can induce a magnon spin accumulation far beyond the nominal magnon spin diffusion length from the injector region. Furthermore, in analyzing the anisotropy in magnon spin transport from the magnitude of the nonlocal second-harmonic resistance, which depends on $|\nabla T|$, is further convoluted with the anisotropy in the thermal conductivity ($\kappa$). Consequently, the observed anisotropy in $R^{\mathrm{NL}}_{2\omega}$ is a combined result of anisotropic heat transport and anisotropy in magnon spin transport. In the case of CrPS$_4$, this interpretation is further supported by Figs. \ref{fig:LocSH} and \ref{fig:NLFH}, where a finite $R^{\mathrm{NL}}_{2\omega}$ (and an anisotropic response along the different axes) is observed even when $R^{\mathrm{NL}}_{1\omega}$ is completely suppressed below the spin-flip field. This observation suggests that $R^{\mathrm{NL}}_{2\omega}$ originates from an extended temperature profile, while the observed anisotropy may arise from differences in the temperature profiles along the two crystallographic axes. The observation of a finite nonlocal second-harmonic resistance even in the absence of a detectable first-harmonic resistance suggests that thermally generated magnons originate from a temperature profile extending well beyond the region defined by the local Joule-heating of the injector. This interpretation is further supported by the local SSE, in Fig. \ref{fig:LocSH}, which exhibits no significant difference up to the spin-flip field and is driven by $|\nabla_c T|$. In contrast, the nonlocal SSE displays a pronounced difference in magnitude even below the spin-flip transition, suggesting that the observed anisotropy originates from differences in both the magnitude of the temperature gradient and the spatial temperature profile along the two crystallographic axes.\\

\begin{figure*}[htb]
  \centering
  \includegraphics[width=\textwidth]{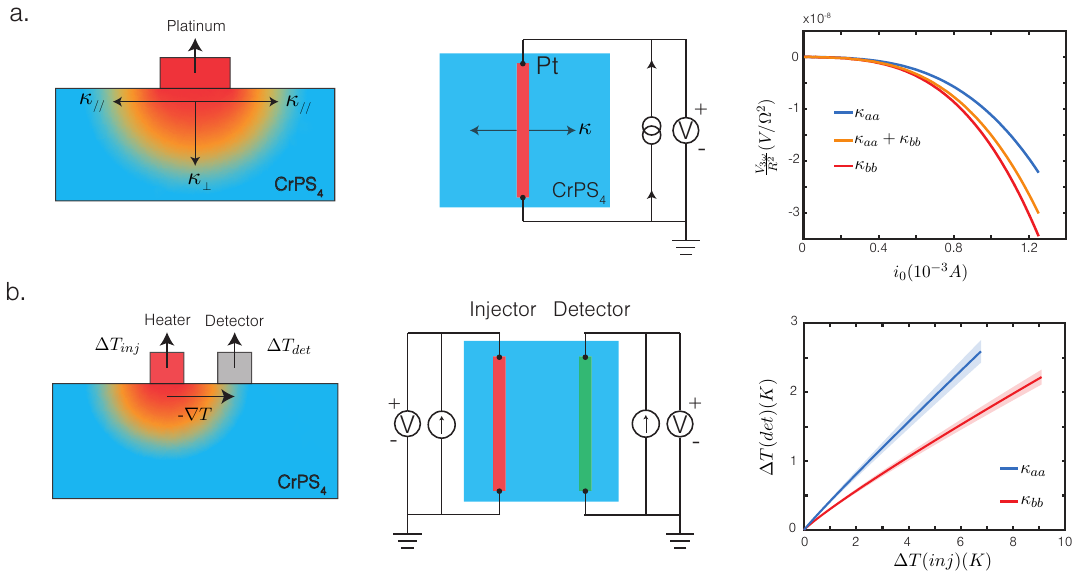}   
    \caption{a) (left) Schematic of the local 3$\omega$ measurement configuration, wherein a single platinum electrode serves as both the heater electrode and the temperature sensor, together with the corresponding temperature profile illustrating heat transport away from the heater, (middle) illustration of the measurement circuit used for the local 3$\omega$ measurements and (right) the local third-harmonic voltage normalized by R$^2$ as a function of $i_0$ for platinum strips oriented along the different crystallographic axes, namely along $a$ ($\kappa_{bb}$), $\frac{a+b}{\sqrt{2}}$ and $b$ ($\kappa_{aa}$) (at 25K); b) (left) Illustration of the nonlocal thermometry configuration, wherein one platinum electrode acts as the heater while a spatially separated platinum strip serves as the temperature sensor, (middle) illustration of the measurement circuit used for the nonlocal thermometry measurements, (right) estimated temperature rise of the nonlocal detector electrode as a function of the temperature rise of the heater electrode at 25 K (the shaded region denotes the uncertainty in the determination of $\Delta T$).}
    \label{fig:Thermometry}
\end{figure*}

The effect of a spatially extended temperature profile is further evident from the distance dependence of the nonlocal first- and second-harmonic resistances shown in Fig. \ref{fig:Distance_dependence}. For the nonlocal first-harmonic resistance, we observe a systematic reduction in the magnitude of the response with increasing injector–detector separation for both crystallographic axes. In contrast, the nonlocal second-harmonic resistance exhibits a clear distance-dependent decrease in magnitude only along the crystallographic \textit{a} axis and only for magnetic fields above the spin-flip transition. For transport along the crystallographic \textit{-b} axis, however, the magnitude of the nonlocal second-harmonic resistance remains nearly unchanged with increasing separation, despite a pronounced decay in the corresponding first-harmonic resistance. For the injector--detector spacings investigated in this work, a similar distance dependence of the nonlocal second-harmonic resistance is observed in devices S2 and S4 (see Supplemental Material XII~\citen{Sup_Inf}), indicating that this behavior is reproducible across the measured devices. \\

Similar to the observations reported in Ref. \citen{Ludo_Nat}, we observe a sign reversal of the nonlocal second-harmonic resistance upon lowering the base temperature, such that the local and nonlocal second-harmonic resistances exhibit the same sign (see Supplemental Material X \citen{Sup_Inf}). This reversal is observed only for small injector--detector separations, further indicating that the mechanisms governing electrically and thermally generated magnon transport are fundamentally distinct. The temperature dependence of the nonlocal second-harmonic resistance additionally highlights the important role of thermal conductivity in thermal magnon generation, particularly since the thermal conductivity decreases significantly at low temperatures altering the temperature profile \citen{Kappa_CPS}.\\

To investigate the possible origins of the anisotropy observed in the nonlocal second-harmonic resistance below the spin-flip field, we examine the anisotropy of heat transport in CrPS$_4$ using thermometry measurements. Specifically, we employ four-probe platinum heaters and perform local $3\omega$ thermometry measurements \citen{euler2015thermal,ThreeOmega_1,ThreeOmega_2}, wherein the platinum injector simultaneously acts as the heater and the temperature sensor as illustrated in Fig. \ref{fig:Thermometry}a. Briefly, for an applied charge current through the platinum heater, Joule heating raises the local temperature of the electrode such that the voltage across the heater is

\begin{equation}
V=iR=i\left(R+\left.\frac{\partial R}{\partial T}\right|_{T=T_0}\Delta T\right),
\end{equation}

where $\Delta T$ is the temperature rise of the heater. Since Joule heating scales with the dissipated power, the temperature rise can be expressed as $\Delta T=\alpha i_0^2R$, where $\alpha$ depends on the thermal transport of CrPS$_4$. The typical four-probe resistance of the Pt electrodes employed for thermometry measurements in this work ranges from $250~\Omega$ to $450~\Omega$. A larger thermal conductivity, $\kappa$, enhances the heat flow away from the heater, thereby reducing the temperature rise and consequently decreasing $\alpha$. Thus, $\alpha$ is inversely related to the thermal conductivity ($\partial\alpha/\partial\kappa<0$), enabling local 3$\omega$ thermometry to probe the relative heat transport along different crystallographic directions in CrPS$_4$. For an applied current $i_{\mathrm{ac}}=\sqrt{2}i_0\sin(\omega t)$, the third-harmonic voltage normalized by the square of the resistance is given by

\begin{equation}
\frac{V_{3\omega}}{R^2}
=
-\left[\frac{1}{2\rho}
\frac{\partial \rho}{\partial T}
\,\alpha\right] i_0^3 ,
\end{equation}

where $\rho$ is the resistivity of platinum. Therefore, the magnitude of the slope of $\frac{V_{3\omega}}{R^2}$ as a function of $i_0^3$ is proportional to $\alpha$. Since all platinum electrodes are deposited simultaneously on the same sample, variations in $\rho$ between electrodes are negligible. Furthermore, $\rho$ is an intrinsic property of platinum, and the four-probe measurement geometry eliminates contributions from the contact resistance. Consequently, differences in the measured slope directly reflect differences in $\alpha$, and hence in the heat transport of the underlying CrPS$_4$.\\

From Fig.~\ref{fig:Thermometry}a (right panel), we find that the magnitude of $\frac{V_{3\omega}}{R^2}$ as a function of $i_0^3$ is largest for platinum strips patterned along the crystallographic \textit{a}-axis of CrPS$_4$ compared to those patterned along the crystallographic \textit{b}-axis. Assuming that the platinum strips act as line heaters, an electrode oriented along the \textit{a}-axis predominantly probes heat transport along the crystallographic \textit{b}- and \textit{c}-directions, whereas an electrode oriented along the \textit{b}-axis is primarily sensitive to heat transport along the \textit{a}- and \textit{c}-directions. Therefore, we attribute the observed differences in $\frac{V_{3\omega}}{R^2}$ as a function of $i_0^3$  to the anisotropy of the in-plane heat conductivity. The larger local temperature rise for the electrode oriented along the \textit{a}-axis indicates less efficient heat transport in the perpendicular \textit{b}-direction, yielding $\kappa_{bb}<\kappa_{aa}$. \\

This conclusion is further supported by the nonlocal thermometry measurements illustrated in Fig.~\ref{fig:Thermometry}b, wherein for a fixed temperature rise of the injector electrode, a larger temperature increase is observed for detector electrodes when the heat transport occurs predominantly along the crystallographic \textit{a}-axis. These measurements likewise indicate that $\kappa_{bb}<\kappa_{aa}$. The device geometries and analysis procedures used to determine the relative thermal conductivities are discussed in detail in Supplemental Material VI~\citen{Sup_Inf}. Furthermore, using the parameters reported in Ref.~\citen{CPS_Mechanical}, obtained from the anisotropic mechanical response of nanomechanical resonators, together with a simplified kinetic theory framework~\citen{LandauLifshitz,DebyeCallaway_DFT,Klemens}, we obtain a qualitative estimate of the thermal conductivity anisotropy in CrPS$_4$, which is consistent with $\kappa_{bb}<\kappa_{aa}$ (see Supplemental Material VI). We emphasize that the thermometry approach employed here determines only the relative thermal conductivity along the different in-plane crystallographic directions and does not provide its absolute value. The absolute thermal conductivity and its temperature dependence along each crystallographic axis in CrPS$_4$ remains to be established. We note that the possible magnetic field dependence of the thermal conductivity of CrPS$_4$ ($\kappa (B)$) is not considered here and remains the subject of ongoing investigation \citen{Krishna}.\\


\section{Conclusion}

The main observations of magnon spin transport along the different crystallographic axes of CrPS$_4$ are summarized below:

\begin{itemize}
    \item For the local SSE response, we exclude contributions from the ordinary Nernst effect. Up to the spin-flip transition, the local second-harmonic resistance, $R^{\mathrm{Loc}}_{2\omega}$, exhibits a similar field dependence for Pt electrodes oriented along the crystallographic \textit{a-} and \textit{b-} axes. This observation suggests that, below the spin-flip field, magnon diffusion along the out-of-plane (\textit{c-}) direction dominates the local SSE response. Above the spin-flip transition, however, the field dependence differs for the two electrode orientations, possibly reflecting the onset of in-plane magnon spin transport. The physical origin of the field dependence of $R^{\mathrm{Loc}}_{2\omega}$ along the different axes above the spin-flip field remains an open question and requires further investigation.

    \item For electrically injected and detected magnons, we observe that the nonlocal first-harmonic resistance, $R^{\mathrm{NL}}_{1\omega}$, arising from magnon transport is larger along the crystallographic \textit{b-} axis of CrPS$_4$ for comparable injector--detector separation. From the distance dependence of $R^{\mathrm{NL}}_{1\omega}$, we resolve the anisotropy in both the magnon spin conductivity and the magnon spin diffusion length, finding them to be larger along the \textit{b-} axis by factors of at least $\sim 2.22$ and $\sim 2.72$, respectively. Along the crystallographic \textit{a-} axis, the magnon spin diffusion length is determined to be approximately 211~nm. In contrast, $R^{\mathrm{NL}}_{1\omega}$ measured along the \textit{b-} axis follows an approximately $1/d$ dependence over the injector--detector separations studied here, indicating that the magnon spin diffusion occurs in the regime where magnon relaxation is negligible. We therefore conclude that $\lambda_m^{b} \geq 575$~nm.  
    
    \item For the nonlocal second-harmonic resistance, $R^{\mathrm{NL}}_{2\omega}$, attributed to the nonlocal SSE, we observe a significantly stronger anisotropy than that in $R^{\mathrm{NL}}_{1\omega}$. The distance dependence of $R^{\mathrm{NL}}_{2\omega}$ reveals that its magnitude decreases with increasing injector--detector separation only for magnon transport along the crystallographic \textit{a-} axis and only above the spin-flip transition. In contrast, $R^{\mathrm{NL}}_{2\omega}$ measured for transport along the crystallographic \textit{b-} axis remains essentially independent of the injector--detector distance in Sample S1, even above the spin-flip field. This behavior differs drastically from that of $R^{\mathrm{NL}}_{1\omega}$, which exhibits a systematic decrease with increasing injector--detector separation. These observations demonstrate that $R^{\mathrm{NL}}_{2\omega}$ reflects a convolution of the spatial temperature profile and magnon spin transport rather than just the latter alone. Consequently, extracting the magnon spin diffusion length directly from $R^{\mathrm{NL}}_{2\omega}$ would substantially overestimate its true value.

    \item From the discussion of the thermally driven magnon transport presented here, we reemphasize that it is therefore crucial to properly account for the full temperature profile within the magnetic insulator when extracting the magnon spin diffusion length from nonlocal SSE. This is particularly important when the extraction relies solely on thermally generated magnon spin currents, in the absence of electrically injected and detected magnons, as in Refs.~\citen{NatCom_Crocl,APL_ThermalMag,xing2019magnon}. For instance, Ref.~\citen{APL_ThermalMag} reported a magnon spin diffusion length of 1.6~$\mu$m for CrPS$_4$ at 20 K, extracted from the nonlocal second-harmonic response. It is furthermore crucial to verify that $R^{\mathrm{NL}}_{2\omega}$ does not exhibit a 1/$d^2$ decay as emphasized in Refs. \citen{Bart_Thermal,Cassanova_Thermal}. Our analysis from the nonlocal first-harmonic response yields $\lambda_m \approx 210$~nm and $\ge 575$~nm along the crystallographic \textit{a}- and \textit{b}-axes, respectively. This substantial discrepancy between the magnon spin diffusion lengths extracted from the nonlocal first- and second-harmonic resistances demonstrates that neglecting the extended temperature profile in the analysis of the nonlocal spin Seebeck effect can lead to a significant overestimation of $\lambda_m$. Accurate determination of the magnon spin diffusion length from $R^{\mathrm{NL}}_{2\omega}$ therefore requires a quantitative treatment of the complete temperature distribution within the magnetic insulator.

    \item Using thermometry measurements we further find that the heat transport in CrPS$_4$ is anisotropic as well with $\kappa_{bb} < \kappa_{aa}$. We note that the anisotropy in the thermal conductivity we observe here is opposite to that calculated in Ref. \citen{NatCom_Aniso}. An important consequence of $\kappa_{bb}<\kappa_{aa}$ is that, for a given temperature rise of the local platinum heater, the resulting thermal gradients are themselves anisotropic, namely that $|\nabla_b T| > |\nabla_a T|$. Since the nonlocal SSE is driven by these thermal gradients, the observed anisotropy in $R^{\mathrm{NL}}_{2\omega}$ is intrinsically intertwined with the anisotropy of the thermal conductivity. We reemphasize that in contrast, the nonlocal first-harmonic resistance, which arises from electrically generated and detected magnons provides a well defined transport channel for extracting both the $\sigma_m$ and $\lambda_m$ along different crystallographic directions. The present results suggest that interpretations of anisotropic $R^{\mathrm{NL}}_{2\omega}$ in terms of anisotropic magnon diffusion alone, as discussed in Refs. \citen{NatCom_Aniso,NatCom_Crocl}, should also take into account the possible influence of anisotropic thermal transport and account for the extended temperature profile in the magnetic material.\\ 

\end{itemize}

In this work, we focus on the in-plane anisotropy of the first-harmonic nonlocal resistance, corresponding to magnon spin transport along the crystallographic \textit{a-} and \textit{b-} directions, while the influence of the thickness of CrPS$_4$ on the magnon spin transport remains the subject of ongoing investigation \citen{Zohaib_InPrep}. The observation of strongly anisotropic magnon spin conductivity and spin relaxation length establishes CrPS$_4$ as a promising platform for investigating anisotropic magnon spin transport phenomena in van der Waals magnets, including the magnon planar Hall-like effects. Furthermore, the demonstration of gate-tunable magnetic properties in CrPS$_4$ \citen{CPS_Electrical}, combined with the anisotropic magnon spin transport demonstrated in this work, opens the possibility of electrically controlling magnon spin transport and tuning the anisotropy of the magnon transport. Although the quantitative role of antiferromagnetic magnon modes in nonlocal magnon transport remains to be clarified, our results provide important insights into how exchange anisotropy governs electrically generated magnon spin transport and influences both local and nonlocal spin Seebeck effects.

\section*{Acknowledgements}

The authors acknowledge the fruitful discussions with J. Barker, M. Mostovoy, G. E. W. Bauer and  Sergio Alvarruiz. The authors acknowledge the technical support from J. G. Holstein, F. H. van der Velde, H. H. de Vries, H. Adema and A. Joshua. The authors acknowledge the research program “Materials for the Quantum Age” (QuMat) for financial support. This program (registration number 024.005.006) is part of the Gravitation program financed by the Dutch Ministry of Education, Culture and Science (OCW). This work is additionally supported by the Zernike Institute for Advanced Materials (ZIAM) at University of Groningen, the Spinoza prize awarded to B.J. van Wees by the Nederlandse Organisatie voor Wetenschappelijk Onderzoek (NWO) in 2016, as well as by the European Research Council (ERC) under the European Union’s 2DMAGSPIN (Grant No. 101053054). 


\bibliographystyle{apsrev4-2}
\bibliography{Reference}

\foreach \p in {1,...,29}{%
  \clearpage
  \includepdf[
    pages={\p},
    pagecommand={},
    turn=false
  ]{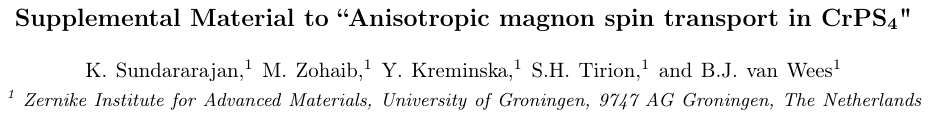}%
}

\end{document}